\documentclass[prc,superscriptaddress,unsortedaddress,twocolumn]{revtex4}
\usepackage{graphicx,epsf,bm,amsmath,color,CJK}
\usepackage{here}
\def\bk{\mbox{\boldmath $k$}}
\def\bq{\mbox{\boldmath $q$}}
\def\bfsigma{\mbox{\boldmath $\sigma$}}
\def\bfSigma{\mbox{\boldmath $\Sigma$}}
\def\bftau{\mbox{\boldmath $\tau$}}
\def\sixj#1#2#3#4#5#6{\begin{Bmatrix}
                    #1 & #2 & #3 \\ #4 & #5 & #6 \end{Bmatrix} }
\def\ninj#1#2#3#4#5#6#7#8#9{\begin{Bmatrix}
                    #1 & #2 & #3 \\ #4 & #5 & #6 \\
                    #7 & #8 & #9  \end{Bmatrix}}
\begin{document}
\title{Single-particle potential of the $\Lambda$ hyperon in nuclear matter with
chiral effective field theory NLO interactions including effects of $YNN$
three-baryon interactions}
\author{M. Kohno}
\email[]{kohno@rcnp.osaka-u.ac.jp}
\affiliation{Research Center for Nuclear Physics, Osaka University, Ibaraki 567-0047, Japan}

\begin{abstract}
Adopting hyperon-nucleon and hyperon-nucleon-nucleon interactions parametrized
in chiral effective field theory, single-particle potentials of the $\Lambda$ and $\Sigma$
hyperons are evaluated in symmetric nuclear matter and in pure neutron matter within
the framework of lowest order Bruckner theory. The chiral NLO interaction bears strong
$\Lambda$N-$\Sigma$N coupling. Although the $\Lambda$ potential is repulsive
if the coupling is switched off, the $\Lambda$N-$\Sigma$N correlation brings about
the attraction consistent with empirical data. The $\Sigma$ potential is repulsive,
which is also consistent with empirical information. The interesting result is that
the $\Lambda$ potential becomes shallower beyond normal density. This provides
the possibility to solve the hyperon puzzle without introducing ad hoc assumptions.
The effects of the $\Lambda$NN-$\Lambda$NN and $\Lambda$NN-$\Sigma$NN
three-baryon forces are considered. These three-baryon forces are first reduced to
normal-ordered effective two-baryon interactions in nuclear matter and then incorporated
in the $G$-matrix equation. The repulsion from the $\Lambda$NN-$\Lambda$NN interaction
is of the order of 5 MeV at the normal density, and becomes larger with increasing the density.
The effects of the $\Lambda$NN-$\Sigma$NN coupling compensate the repulsion at normal
density. The net effect of the three-baryon interactions to the $\Lambda$ single-particle
potential is repulsive at higher densities.  
\end{abstract}

\maketitle

\section{Introduction}
It is a fundamental problem to understand hyperon properties in the hadronic medium
based on underlying baryon-baryon interactions. Despite the scarceness of
hyperon-nucleon (YN) scattering data, various YN potential models have been developed
by several groups \cite{NIJM,NSC97,NIJ89,JUE05,KN07}. The empirical data
of $\Lambda$ hyper-nuclei have suggested that the depth of the $\Lambda$
single-particle (s.p) potential in the nuclear medium is about 30 MeV \cite{MOT88,HAS06}.
Most of the YN potentials account for the $\Lambda$-nucleus attractive interaction of
this order, for example, in the framework of Brueckner theory \cite{YB90,SCH98,KOH00,VID00}.
Calculations for neutron-star matter at higher densities with those interactions have shown
that the $\Lambda$ hyperons become energetically favored to bypass the increasing
neutron Fermi energy at $2\sim 3$ $\rho_0$, where $\rho_0=\frac{2k_F^3}{3}=0.166$ fm$^{-3}$
with the Fermi momentum $k_F=1.35$ fm$^{-1}$ is referred to as normal density.
Beyond the onset of the $\Lambda$ emergence, the equation of state (EOS) of high-density
nuclear matter naturally becomes soft. Such an EOS is difficult to explain even the standard
neutron-star mass of $1.4$ solar mass ($1.4 M_\odot$) \cite{NIS01,SCH06}.
Recently, the situation has become more serious, after neutron stars with a mass of
$2 M_\odot$ were observed \cite{DEM10,ANT13}. The problem is called hyperon puzzle.
It is necessary to advance the study of the interaction between the $\Lambda$ hyperon
and the nucleons in the nuclear medium.

In the non-strangeness sector, the potentials derived in the framework of chiral effective
field theory (ChEFT) \cite{MACH,EHM09} have been widely employed in recent ab initio
calculations of properties of atomic nuclei on the basis of nucleon-nucleon interaction.
In addition to their comparable accuracy in describing nucleon-nucleon (NN) scattering data
with other modern NN potentials, three-nucleon forces (3NFs) are introduced systematically
and consistently with the NN sector. Baryon-baryon interactions in the strangeness
sector have also been developed, although the experimental data are still limited.
The parameterization of the YN potential in the lowest order of the chiral expansion was
given by Polinder \textit{et al.} \cite{POL06}. The extension to the next-to-leading order (NLO)
was achieved by Haidenbauer \textit{et al.} \cite{HAID13}. Pion-exchange YNN three-baryon
forces (3BFs) were recently derived by Petschauer \textit{et al.} \cite{PET16}.

In this article, $\Lambda$ and $\Sigma$ single-particle (s.p.) potentials are investigated,
in the framework of lowest-order Brueckner theory (LOBT), in symmetric nuclear matter
(SNM) and in pure neutron matter (PNM), using the ChEFT NLO interactions.
The density-dependence of the $\Lambda$ s.p. potential in pure neutron matter is
relevant to the role of strangeness in neutron star matter.
Because ChEFT is low-momentum effective theory with the cutoff scale of
about 500 MeV, it is not applicable beyond the density of the $2\sim2.5 \rho_0$.
Nevertheless, it is worth to investigate properties of the $\Lambda$ hyperon in
nuclear matter predicted by the microscopic ChEFT YN interactions below the
limit as the basis towards the higher densities.

3NFs are indispensable to properly describe the saturation properties of nuclear matter.
Therefore, it is important to estimate the effects of YNN 3BFs for hyperon
properties in the nuclear medium. The advantage of ChEFT is
that the 3BFs are systematically introduced. In this article, two-pion exchange
$\Lambda$NN-$\Lambda$NN and $\Lambda$NN-$\Sigma$NN 3BFs are considered,
following Petschauer \textit{et al.} \cite{PET16}. The contributions of the 3BFs are
evaluated by introducing density-dependent effective two-body YN interactions
by integrating one nucleon degrees of freedom in the medium.
The procedure may be referred to as normal-ordered prescription in the medium.

Hyperon properties calculated by the ChEFT interactions have been reported by
Haidenbauer and Mei{\ss}ner \cite{HM15} and by Petschauer \textit{et al.} \cite{EPJ16}.
The recent publication by Haidenbauer \textit{et al.} \cite{HMKW17} discusses the
implication of the ChEFT interactions to the hyperon puzzle. The present article
presents the results from independent calculations, which are qualitatively similar
to the preceding calculations as it should because the same ChEFT hyperon-nucleon
interactions are employed. The treatment of the 3BFs is improved to use
general expressions for the off-shell components. The effects of the
$\Lambda$NN-$\Sigma$NN are taken into account, which have not been
considered before.

In section II, $\Lambda$ and $\Sigma$ LOBT s.p. potentials are evaluated at
several densities first in SNM and next in PNM, using the NLO YN interactions only.
The depth of the $\Lambda$ potential at the normal density is seen to be consistent
with the empirical value. Salient features of the results are discussed.
The outline of evaluating the normal-ordered YN interactions from the 3BFs is given
in Sec. III. The $\Lambda$NN 3BF is included first, and then the influence of the
$\Lambda$NN-$\Sigma$NN coupling interactions is incorporated.
Explicit expressions of the normal-ordered YN interactions are presented in Appendix.
The numerical results of the 3BF contributions both in SNM and
PNM are shown in Sec. IV. Summary follows in Sec. V.

\section{Hyperon properties with NLO YN interactions}
\subsection{symmetric nuclear matter}
Hyperon-nucleon interactions derived within ChEFT are not soft enough to be
directly used in tha mean field description or perturbative method especially for the
$\Lambda$N-$\Sigma$N coupling. It is necessary to introduce some effective
interactions appropriate to lower-energy scale. The G-matrix equation in
conventional Brueckner theory takes care of in-medium correlations to treat
short-range repulsive part, and medium effects such as Pauli blocking as well as
dispersion effects. The Brueckner self-consistency for the single-particle
potentials amounts to including a certain set of higher-order diagrams.

Because the lowest-order Bruckner calculations have been extensively carried out
in the literature \cite{YB90,SCH98,KOH00,VID00}, it is sufficient to note some
comments for the present application as follows.
1) The continuous choice is used for the intermediate spectra.
An effective mass approximation is not used, but s.p. energies are interpolated
from the values at the mesh points.
2) The angle-average approximation is introduced for the Pauli operator in the
numerator and the energies in the denominator of the propagator.
3) Partial waves up to the total angular momentum $J=6$ are included.

Before carrying out G-matrix calculations for hyperons, it is necessary to prepare
nucleon single-practice energies, which are needed for the propagator in the
G-matrix equation. In the present calculations, the nuclear matter properties
\cite{MK13} obtained with the potential parametrized at the N$^3$LO level by
the Bochum-Bonn-J{\"{u}lich group \cite{EGM05} are used, which include
the effects of the leading-order 3NF as the density-dependent effective interactions
by folding the third nucleon degrees of freedom. The components of the 3NF
which is determined by the coupling constants fixed in the NN sector provide
strong repulsion. This repulsion is decisively important to account for nuclear
saturation properties, because the minimum of the saturation curve obtained
by any realistic NN interaction locates at considerably higher densities than
the empirical one. The coupling constants in the one-pion exchange contact and
three-nucleon contact terms, respectively, are tuned to reasonably reproduce
a saturation curve as $c_D=-2.5$ and $c_E=0.25$ for the case of the cutoff
$\Lambda=550$ MeV \cite{MK13}.

\begin{figure}[tbh]
\centering
 \includegraphics[width=0.4\textwidth,clip]{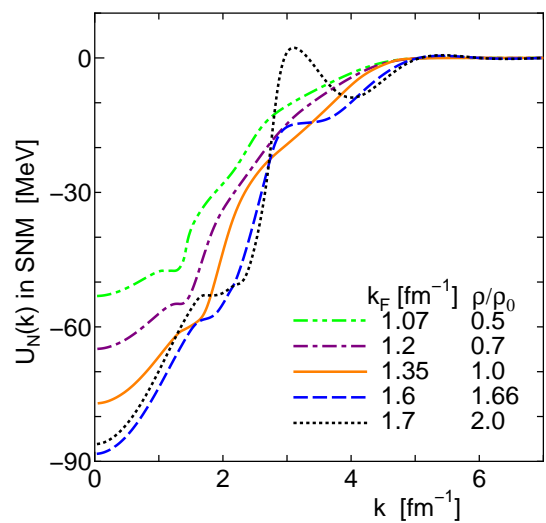}
 \caption{Nucleon s.p. potentials as a function of the nucleon momentum $k$
obtained from the LOBT calculations in SNM with the ChEFT NN potential
\cite{EGM05} of the cutoff $\Lambda=550$ MeV, including the effects from
the 3BF. These potentials are used for the propagator in the YN $G$-matrix
equation.}
\label{fig:SNMsp}
\end{figure}

The s.p. potential at the large momentum, typically $k> 3$ fm$^{-1}$, starts
to show oscillation as an artifact of the cutoff, when the Fermi momentum $k_F$
becomes large. This behavior makes it difficult to numerically obtain
a Brueckner self-consistent solution. In the present calculations, a cutoff
regularization in the form of $e^{-(k^2/(2\Lambda^2))^3}$
is applied to the single particle potential used in the $G$-matrix equation.
It has been checked that the s.p. potential in the lower momentum region,
$k<2$ fm$^{-1}$, has scarcely been altered by this procedure.
The same prescription is applied also for the hyperon s.p. potentials.
The momentum dependence of the nucleon s.p. potential with the above
regularization is shown in Fig. \ref{fig:SNMsp} for SNM and in Fig. \ref{fig:PNMsp}
for PNM. The nucleon s.p. potential does not become monotonically deeper as
the density becomes larger because of the repulsive contribution of the 3NFs. 

\begin{figure}[tbh]
\centering
 \includegraphics[width=0.4\textwidth,clip]{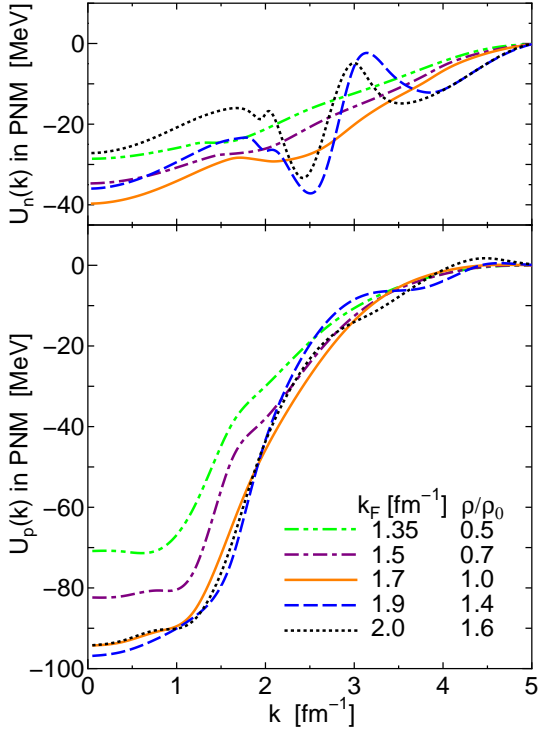}
 \caption{Same as Fig. \ref{fig:SNMsp}, but for neutron and proton s.p. potentials
in PNM.}
\label{fig:PNMsp}
\end{figure}

Calculated results for the $\Lambda$ and $\Sigma$ s.p. potentials in SNM with
the chiral NLO YN potentials \cite{HAID13} are shown in Fig. \ref{fig:SNMR0}
for the real part and in Fig. \ref{fig:SNMI0} for the imaginary part, for the four
values of the Fermi momentum; $k_F=1.07$, 1.35, 1.6 and 1.7 fm$^{-1}$.
The corresponding densities of nuclear matter are $0.50\rho_0$, $\rho_0$,
$1.66\rho_0$, and $2.0\rho_0$, respectively. Solid curves represent the results
of the calculation in which the $\Lambda$N-$\Sigma$N coupling is normally
included. The dotted curves are the potential after the regularization factor
$e^{-(k^2/(2\Lambda^2))^3}$ is multiplied, which is employed in the denominator of
the propagator of the $G$-matrix equation. To assure that the introduction of the
regularization factor for the sake of numerical stability does not change low-energy
quantities, the results without applying this prescription are shown by a dashed curve
for $k_F=1.35$ and $1.60$ fm$^{-1}$. The solid and dashed curves do not differ in
a low-momentum region.

The importance of the $\Lambda$N-$\Sigma$N coupling, to which the tensor force
from the one-pion exchange contributes, has been recognized in every realistic
YN potential based on the underlying picture of meson exchanges.
To quantify the effect of this coupling, the $\Lambda$ and $\Sigma$ s.p. potentials
are evaluated by switching off the coupling, the results of which are indicated by
the dot-dashed curves in Fig. \ref{fig:SNMR0}. The effect is seen to be particularly
sizable. The $\Lambda$ s.p. potential without the $\Lambda$N-$\Sigma$N coupling
is even repulsive. The potential depth of about $30$ MeV at the normal density,
which is consistent with the empirical

\begin{figure}[H]
\centering
 \includegraphics[width=0.4\textwidth,clip]{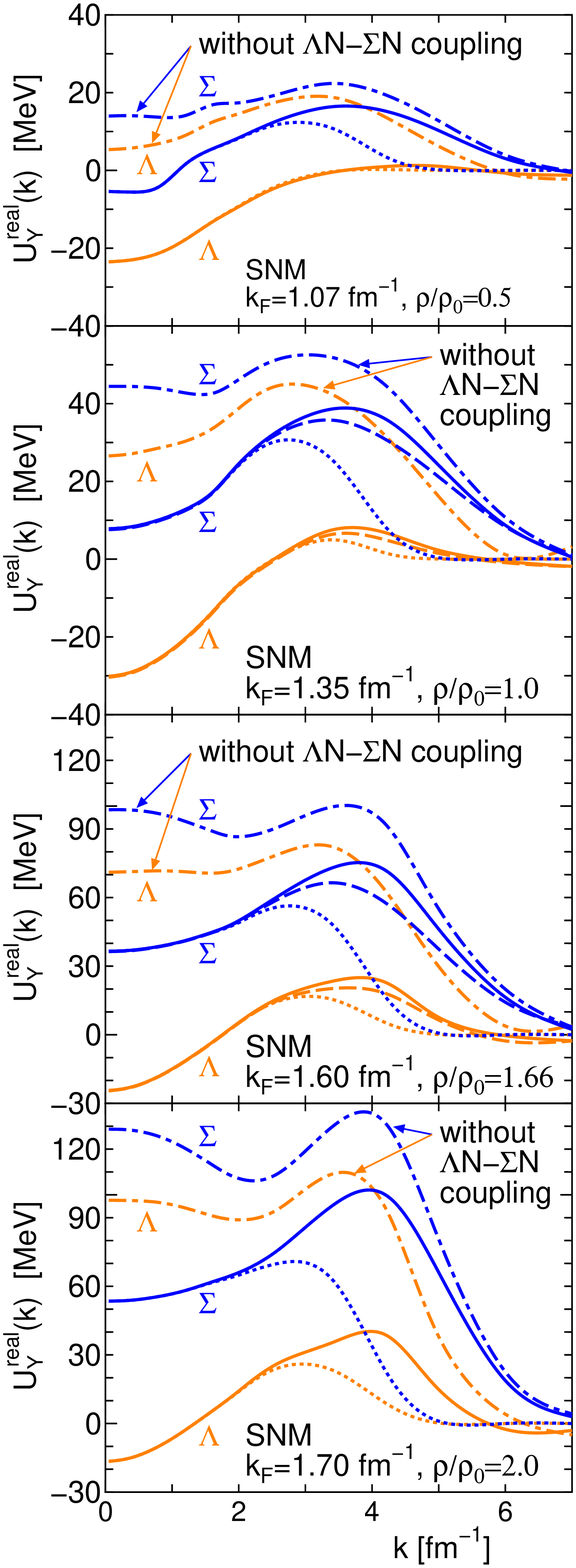}
 \caption{Real part of the $\Lambda$ and $\Sigma$ s.p. potentials in SNM
with the chiral NLO YN interactions \cite{HAID13}. The solid and dot-dashed
curves represent the results of the calculation with switching on and off
the $\Lambda$N-$\Sigma$N coupling, respectively.
The potentials multiplied by the regularization factor $e^{-(k^2/(2\Lambda^2))^3}$
with $\Lambda=550$ MeV are shown by the dotted curves.
The results of the calculation in which the prescription of the regularization is
not applied are shown by the dashed curves for $k_F=1.35$ and $1.60$ fm$^{-1}$.}
\label{fig:SNMR0}
\end{figure}

\begin{figure}[thb]
\centering
 \includegraphics[width=0.4\textwidth,clip]{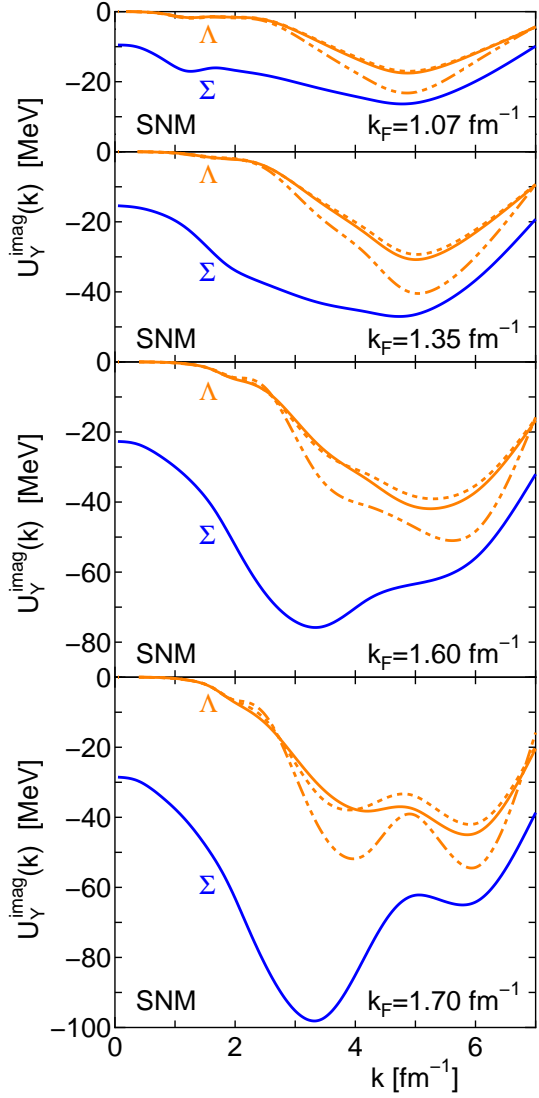}
 \caption{Imaginary part of the $\Lambda$ and $\Sigma$ s.p. potentials in SNM
with the chiral NLO YN interactions \cite{HAID13}.
The short dashed and two-dot-dashed curves stand for
the results including the effects from $\Lambda$NN-$\Lambda$NN 3BFs and
$\Lambda$NN-$\Sigma$NN 3BFs in addition, respectively, which are discussed
in Sec. III.}
\label{fig:SNMI0}
\end{figure}

\noindent
data \cite{MOT88,HAS06}, is
brought about by the attraction from the coupling. Though the coupling also
yields the attraction for the $\Sigma$ hyperon, the $\Sigma$ potential is still positive
at the normal density. At low densities, e.g. at $\rho/\rho_0=0.5$ in Fig. \ref{fig:SNMR0},
the $\Sigma$ s.p. potential becomes attractive. These features are also
consistent with experimental information \cite{SAHA04,MK06}.

The partial-wave contributions to the s.p. potential of $\Lambda$ at rest, which
are shown in Fig. \ref{fig:lampw}, detail the properties of the NLO $\Lambda$N interaction
and the $\Lambda$N-$\Sigma$N coupling. The solid and dashed curves are the results
with switching on and off the $\Lambda$N-$\Sigma$N coupling, respectively.
The characteristic feature is the density dependence of the $^3$S$_1$ contribution
when the $\Lambda$N-$\Sigma$N coupling is not taken into account.
The repulsion rapidly grows with increasing the density. Although the
$\Lambda$N-$\Sigma$N coupling brings about sizable attraction, the attractive
$\Lambda$ s.p. potential does not become deeper at greater densities than normal.  
The attractive contribution in the $^1$S$_0$ state is of the similar order for the
$^3$S$_1$ contribution, whereas the contributions from other
channels are small and do not depend much on the density.

The small splitting of the $^3$P$_1$ and $^3$P$_2$ contributions in Fig. 5, which is about
one third of that for the case of the nucleon, indicates that the $\Lambda$N spin-orbit
interaction is not large. However, the experiments have suggested \cite{HAS06} that the
$\Lambda$N effective spin-orbit interaction is very small. In the analysis of the
quark-model $\Lambda$N interaction, it was demonstrated \cite{FUJ00} that the
antisymmetric spin-orbit component cancels the normal spin-orbit contribution to
realize small spin-orbit splitting. In the present NLO $\Lambda$N interaction,
there is no antisymmetric spin-orbit component. The possible reduction of the effective
$\Lambda$ spin-orbit field by incorporating the contact terms at NLO was discussed
by Haidenbauer and Mei{\ss}ner \cite{HM15}. It is noted that the 3BFs considered in
the next section have normal and antisymmetric spin-orbit components, and they reduce
the $\Lambda$N spin-orbit strength by about 10\%. 

The interesting consequence of these properties of the ChEFT YN interactions
is that the depth of the $\Lambda$ s.p. potential does not become deep with
increasing the density.
This behavior is paved by the density-dependence of the $^3$S$_1$ contribution
before taking into account the $\Lambda$N-$\Sigma$N coupling. In addition,
the $\Lambda$N-$\Sigma$N coupling tends to be suppressed in the nuclear
medium with large Fermi momentum by the Pauli blocking for the intermediate
nucleon state. The similar behavior is also seen in PNM. The implication of this
variation of the $\Lambda$ s.p. potential with respect to the density to the
hyperon puzzle is discussed in Sec. IV after including the 3BF effects.

\begin{figure}[t]
\centering
 \includegraphics[width=0.4\textwidth,clip]{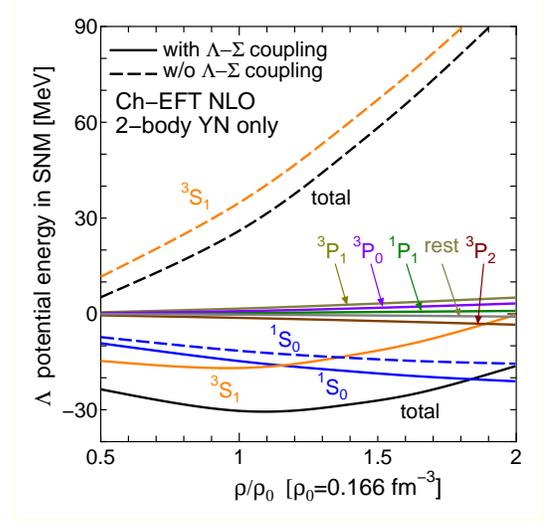}
 \caption{Density-dependence of the partial wave contributions to the $\Lambda$
s.p. potential in SNM with NLO YN interactions \cite{HAID13} only. 
}
\label{fig:lampw}
\end{figure}

The imaginary part of the hyperon s.p. potential is related to the spreading
width of the hyperon s.p. state in the nuclear medium by $\Gamma =-2 \Im U_Y$,
The vary small width of $\Lambda$ hypernuclear states, in comparison with
the nucleon states, has been observed experimentally \cite{HAS06}.
The reason based on the properties of the YN interaction
was discussed in details by Bando \textit{et al.} \cite{BMY85}.
Figure \ref{fig:SNMI0} shows that the imaginary part of the $\Lambda$ s.p.
potential obtained by the ChEFT NLO YN interaction is actually very small,
as in other calculations using different YN potentials \cite{SCH98,KF09}.
The strong $\Lambda$N-$\Sigma$N coupling does not directly contribute
to the imaginary potential. The imaginary $\Sigma$ potential of $\Im U_\Sigma (0)
=\sim -15$ to $\sim -20$ MeV is also similar to those with other
YN potentials \cite{SCH98,KF09}.

\subsection{Pure neutron matter}
In pure neutron matter, the hyperon potential is charge-dependent, and $G$-matrix
equations are solved in a particle-base. Because the $\Sigma^-$ hyperon
does not couple with the $\Lambda$ hyperon, the $\Sigma^-$n $G$-matrix is
calculated by a single-channel equation. The $\Lambda$ and $\Sigma^0$
states are determined through a $\Lambda$n-$\Sigma_0$n-$\Sigma^-$p coupled
equation. The $\Sigma^+$ s.p. potential is obtained from a
$\Sigma^+$n-$\Lambda$p-$\Sigma^0$p coupled equation. These
$\Lambda$ and $\Sigma^{\pm,0}$ s.p. potentials are determined self-consistently.

The calculated results for $k_F^n=1.35, 1.7$, and 2.0 fm$^{-1}$, the neutron densities being
$\frac{1}{2}\rho_0$, $\rho_0$, and $1.63\rho_0$, respectively, are shown
in Fig. \ref{fig:PNM0}. Note that a stable solution was not obtained
for the density beyond $k_F=2.0$ fm$^{-1}$ or $1.6\rho_0$.
The dashed curves for $\Lambda$, $\Sigma^0$, and $\Sigma^+$ represent the results
in which the $\Lambda$N-$\Sigma$N coupling is switched off. With increasing the density
from $k_F^n=1.35$ to 1.7 fm$^{-1}$, the $\Lambda$ s.p. potential at $k=0$ decreases
to about $-30$ MeV, and then it turns to become shallower at higher densities, as in SNM.

Because $\Sigma^-$ does not couple to $\Lambda$ in PNM, the $\Sigma^-$
s.p. potential is resembling to the dot-dashed curve in Fig. \ref{fig:SNMR0}; that is,
repulsive even at low densities.
When the $\Sigma$N-$\Lambda$N coupling is ignored, the ordering of the $\Sigma$
s.p. potential is $0 < U_{\Sigma^-}(0) < U_{\Sigma^0}(0) < U_{\Sigma^+}(0)$.
The $\Sigma^+$ s.p. potential is much influenced by the coupling, and the ordering
is reversed: $U_{\Sigma^+}(0) < 0 < U_{\Sigma^0}(0) < U_{\Sigma^-}(0)$.
Still, the $\Sigma^0$ s.p. potential is considerably repulsive.
The increasingly repulsive nature of the $\Sigma^-$ and $\Sigma^0$ s.p. potentials
with growing the density in PNM suggests that the onset of the emergence of
these $\Sigma^-$ and $\Sigma^0$ hyperons in neutron star matter tends to
be prevented.

\begin{figure}[thb]
\centering
 \includegraphics[width=0.4\textwidth,clip]{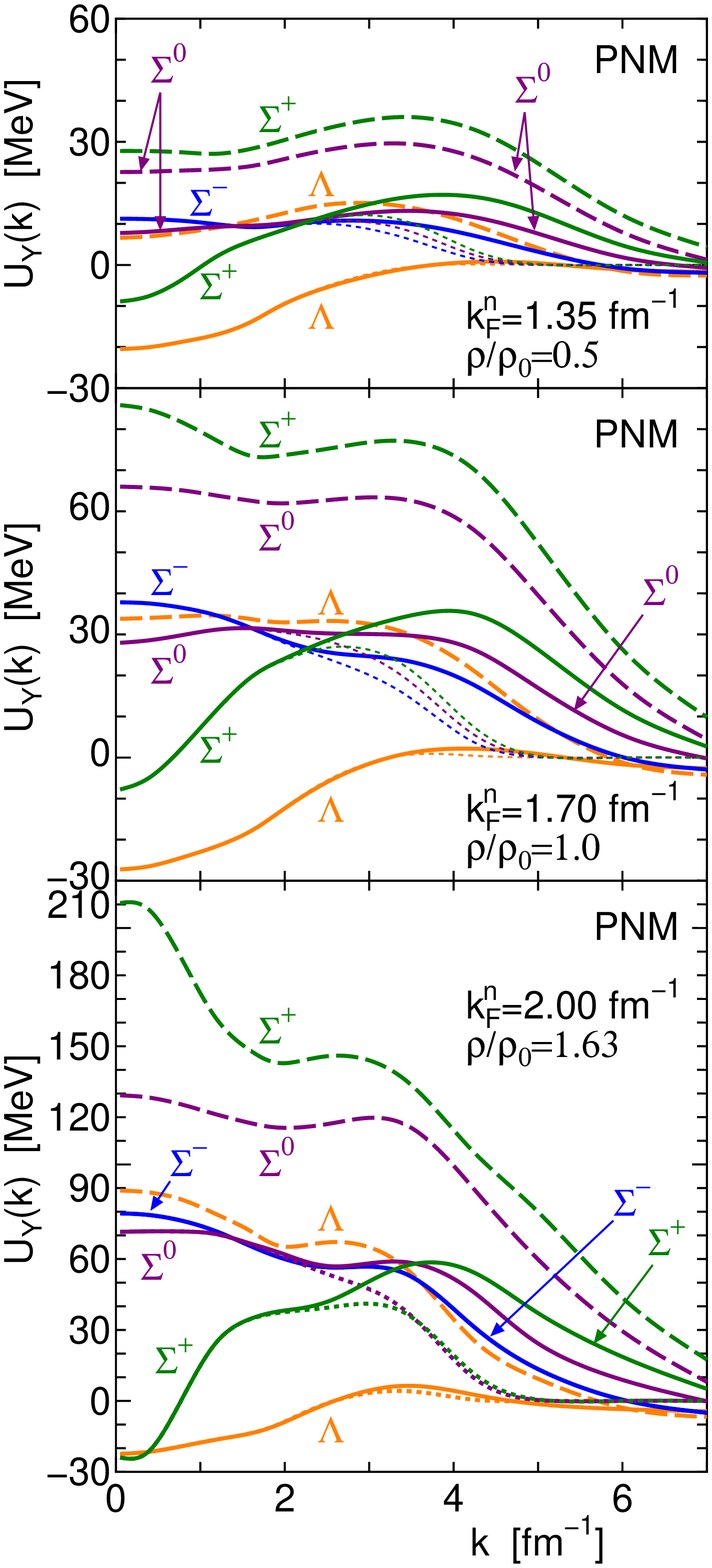}
 \caption{$\Lambda$ and $\Sigma$ s.p. potentials in PNM with the chiral
NLO YN interactions \cite{HAID13}. Solid curves represent the results including the
cutoff form factor for the intermediate potentials in the denominator of the $G$-matrix
equation. The latter potentials are shown by the dotted curves. Calculated results
with switching off the $\Lambda$N-$\Sigma$N coupling are shown by the dashed
curves.}
\label{fig:PNM0}
\end{figure}

\section{Including effects of YNN interactions}
Leading order three-baryon forces (3BFs) were derived by Petschauer \textit{et al.}
\cite{PET16} in SU(3) chiral effective field theory. A feasible way of investigating the
effects of these 3BFs to the $\Lambda$ s.p. potential in nuclear
matter is to reduce them to density dependent effective
two-body interactions by integrating the third nucleon over the occupied states,
which may be called as a normal ordered two-body interaction of the original 3BF
with respect to nuclear matter. In this article, two-pion exchange
$\Lambda$NN-$\Lambda$NN and $\Lambda$NN-$\Sigma$NN 3BFs are taken
into consideration, but the $\Sigma$NN-$\Sigma$NN 3BF is left out, because
its contribution to the $\Lambda$ s.p. potential is through the correction
for the $\Sigma$ s.p. potential in the propagator of the $G$-matrix equation
and therefore indirect.

Petschauer \textit{et al.} \cite{PET17} presented the density-dependent effective
two-body interactions of the 3BFs in the case of $|\bk|=|\bk'|$ in momentum space,
where $\bk$ and $\bk'$ are initial and final relative momenta of the two interacting
baryons. Here, the expressions for the general case of $|\bk| \ne |\bk'|$ are
derived, as in the case of three-nucleon forces \cite{MK13}.

\subsection{Density-dependent $\Lambda$N-$\Lambda$N interaction from
NNLO 2$\pi$-exchange $\Lambda$NN-$\Lambda$NN interaction}
\begin{figure}
\centering
 \includegraphics[width=0.4\textwidth,clip]{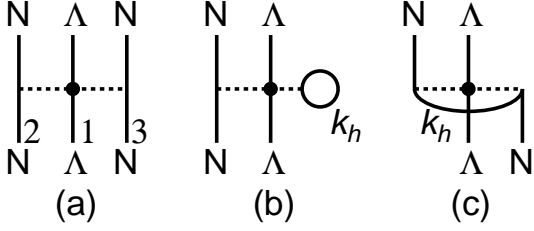}
 \caption{$\Lambda$NN-$\Lambda$NN diagrams. The dotted line represents pion
exchange. (a) $\Lambda$NN-$\Lambda$NN three-body interaction
$V_{TPE}^{\Lambda NN}$, (b) direct term in one-nucleon
folding of $V_{TPE}^{\Lambda NN}$, and (c) exchange term in one-nucleon
folding of $V_{TPE}^{\Lambda NN}$. $\bk_h$ indicates an occupied state.}
\label{LLdiag}
\end{figure}

Because there is no $\Lambda\Lambda\pi$ coupling, the two-pion exchange
$\Lambda$NN-$\Lambda$NN interaction $V_{TPE}^{\Lambda NN}$ is
given by the diagram of Fig. \ref{LLdiag}(a) alone. Following the expression
by Petschauer \textit{et al.} \cite{PET16}, the Born amplitude of this diagram
is written as
\begin{align}
 V_{TPE}^{\Lambda NN} &= \frac{g_A^2}{3f_0^4}
\frac{(\bfsigma_3 \cdot \bq_{3})(\bfsigma_2 \cdot \bq_{2})}
{(\bq_{3}^2+m_\pi^2)(\bq_{2}^2+m_\pi^2)} (\bftau_2\cdot\bftau_3) \notag \\
 & \times \{-(3b_0+b_D)m_\pi^2 +(2b_2+3b_4)\bq_{3}\cdot\bq_{2}\},
\end{align}
where the coordinate 1 is assigned to the $\Lambda$ hyperon and $\bq_{2}$ ($\bq_{3}$)
is the difference of the final and initial momenta at the nucleon line 2 (line 3).
$g_A$ is the axial coupling constant, $f_0$ is the pion decay constant, $m_\pi$
is the pion mass, and $\bfsigma$ and $\bftau$ stand for the spin and isospin operators.
The coupling constants $b_0$, $b_D$, $b_2$, and $b_4$ inherit those in the
underlying Lagrangian. This 3BF is reduced to an effective
$\Lambda$N-$\Lambda$N interaction $V_{TPE}^{\Lambda N(N)}$ by folding
one nucleon degrees of freedom over the occupied states in nuclear matter.
Assuming that two baryons are in the center of mass frame, the effective
two-body matrix element from $V_{TPE}^{\Lambda NN}$ is obtained as
\begin{align}
 & \langle \bk' \sigma_\Lambda', -\bk' \sigma' \tau' |V_{\Lambda N(N)}|
\bk \sigma_\Lambda, -\bk \sigma \tau \rangle \notag \\
 &=  \frac{g_A^2}{3f_0^4} \sum_{\bk_h}\sum_{\sigma_h, \tau_h}
 \langle \bk' \sigma_\Lambda, -\bk' \sigma' \tau', \bk_h \sigma_h \tau_h|
  (\bftau_2\cdot\bftau_3)\notag \\
 & \times \frac{(\bfsigma_3 \cdot (-\bk'-\bk_h)) (\bfsigma_2 \cdot (\bk_h+\bk))}
{((\bk_h+\bk')^2+m_\pi^2)((\bk_h+\bk)^2+m_\pi^2)} \nonumber \\
 & \times\{-(3b_0+b_D)m_\pi^2 +(2b_2+3b_4) (-\bk_h-\bk')\cdot(\bk_h+\bk)\}\notag \\
& \times | (\bk \sigma_\Lambda, -\bk \sigma \tau, \bk_h \sigma_h \tau_h)
-(\bk \sigma_\Lambda, \bk_h \sigma_h \tau_h, -\bk \sigma \tau) \rangle,
\label{eq:LN}
\end{align}
where $\bk_h$, $\sigma_h$ and $\tau_h$ specify the nucleon in momentum, spin, and
isospin states. In the following, $\tau=-1/2$ is assigned for neutron and $\tau=1/2$
proton. The diagrammatic representation of the integration is depicted in
Figs. \ref{LLdiag}(b) and \ref{LLdiag}(c).
The direct term, Fig. \ref{LLdiag}(b), vanishes because of the spin summation.
The isospin summation in the exchange contribution gives
$\sum_{\tau_h} \langle \tau' \tau_h| (\bftau_2\cdot\bftau_3)| \tau_h \tau\rangle
= 3\delta_{\tau'\tau}$ in SNM, and
$\{1\delta_{\tau' -1/2}\delta_{\tau, -1/2}+ 2\delta_{\tau', 1/2} \delta_{\tau,1/2}\}$
in PNM.
While central, spin-orbit, and antisymmetric spin-orbit components appear from
$V_{TPE}^{\Lambda NN}$, no tensor component is generated.
Explicit expressions after carrying out the spin-summation
and $\bk_h$ integration are given in Appendix. If the condition of $|\bk'|=|\bk|$
is imposed for the expression in Appendix, the results agree with those given
by Petschauer \textit{et al.} \cite{PET17}. It should be noted that a statistical
factor of $\frac{1}{2}$ has to be multiplied when the effective two-body interaction
defined by Eq. \ref{eq:LN} is added to the original two-body
$\Lambda$N-$\Lambda$N interaction. It is worthwhile to mention that in the case
of three-nucleon forces the statistical factor is $\frac{1}{3}$ for the total energy
in the Hartree-Fock level and $\frac{1}{2}$ for the s.p. energy, whereas here the
factor of $\frac{1}{2}$ is common for the total and s.p. energies.

As for a form factor, it is not included in the stage of the integration of Eq. \ref{eq:LN},
but the resulting effective two-body interaction $V_{\Lambda N(N)}$ is multiplied
by the following Gaussian regularization factor same as in the NN sector
with the cutoff scale $\Lambda$ being 550 MeV:
\begin{equation}
 \exp(-(k'/\Lambda)^6-(k/\Lambda)^6).
\end{equation}

The normal-ordered interaction $V_{\Lambda N(N)}$ can be regarded as a
Pauli-blocking effect for the two-pion exchange two-body $\Lambda$N
interaction with the $\pi\pi\Lambda\Lambda$ vertex. Supposing that this two-pion
exchange $\Lambda$N interaction provides an attractive component,
the Pauli-blocking brings about repulsion.
 
\subsection{Density-dependent $\Lambda$N-$\Sigma$N interaction from
NLO $\Lambda$NN-$\Sigma$NN interaction}
\begin{figure}[thb]
\centering
 \includegraphics[width=0.45\textwidth,clip]{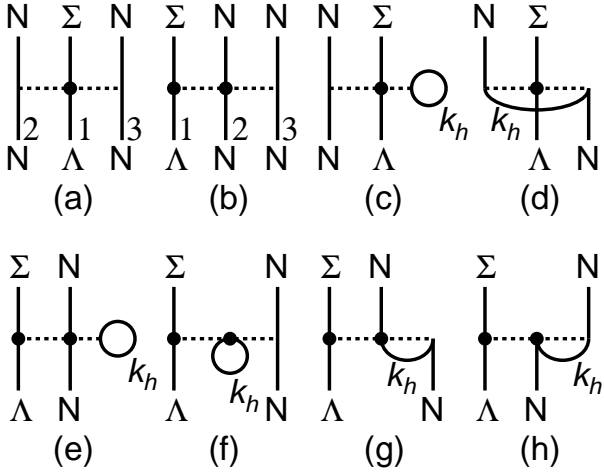}
 \caption{$\Lambda$NN-$\Sigma$NN diagrams. The dotted line represents pion
exchange. (a) 3BF $V_{TPE,a}^{\Lambda-\Sigma}$ with $\pi\pi\Lambda\Sigma$
vertex, (b) 3BF $V_{TPE,b}^{\Lambda-\Sigma}$ with $\pi\Lambda\Sigma$ vertex,
(c) direct term in one-nucleon folding of $V_{TPE,a}^{\Lambda-\Sigma}$,
and (d) exchange term in one-nucleon folding of $V_{TPE,a}^{\Lambda-\Sigma}$,
(e) direct term in one-nucleon folding of $V_{TPE,b}^{\Lambda-\Sigma}$,
(f) exchange term in one-nucleon folding of $V_{TPE,b}^{\Lambda-\Sigma}$,
and (g)  exchange term in one-nucleon folding of $V_{TPE,a}^{\Lambda-\Sigma}$.
$\bk_h$ indicates an occupied state.}
\label{LSdiag}
\end{figure}

As for the $\Lambda$NN-$\Sigma$NN transition process, in addition to the diagram
Fig. \ref{LSdiag}(a), there appears another type of the diagram presented
in Fig. \ref{LSdiag}(b) . Besides, the amplitude of the diagram of Fig. \ref{LSdiag}(a)
has an extra term;
\begin{align}
 V_{TPE,a}^{\Lambda -\Sigma} &= -\frac{1}{4 f_0^4}
\frac{(\bfsigma_2 \cdot \bq_{2})(\bfsigma_3 \cdot \bq_{3})}
{(\bq_{2}^2+m_\pi^2)(\bq_{3}^2+m_\pi^2)} i(\bfSigma\cdot (\bftau_2 \times\bftau_3))\notag \\
 & \times[ N_1^a+N_2^a (\bq_2\cdot\bq_3)
 + N_3^a  i(\bfsigma_1 \cdot (\bq_{2}\times \bq_{3})].
\label{eq:LSLCa}
\end{align}
The amplitude of the diagram of Fig. \ref{LSdiag}(b) has the following structure;
\begin{align}
 V_{TPE,b}^{\Lambda -\Sigma} &= -\frac{1}{4 f_0^4}
\frac{(\bfsigma_1 \cdot \bq_{1})(\bfsigma_3 \cdot \bq_{3})}
{(\bq_{1}^2+m_\pi^2)(\bq_{3}^2+m_\pi^2)} \notag \\
 & \times[ (\bfSigma\cdot \bftau_3)\{N_1^b +N_2^b (\bq_1\cdot\bq_3)\} \notag \\
 & - N_3^b ((\bfSigma\times \bftau_3)\cdot \bftau_2)((\bq_{1}\times \bq_{3})\cdot \bq_2)].
\label{eq:LSLCb}
\end{align}
In the above expressions, $\bfSigma$ represents the isospin operator asociated
with the $\Lambda$-$\Sigma$ transition, and the explicit values of the coupling
constants $N_{1\sim 3}^a$ and $N_{1\sim 3}^b$ are given in the next subsection.

To obtain effective two-body interactions from these 3BFs,
the following matrix element is calculated in which both the bra and ket states are
anti-symmetrized with respect to the two nucleons.
\begin{align}
 & \langle \bk' \sigma_\Sigma \tau_\Sigma, -\bk' \sigma' \tau' |V_{\Lambda N(N)-\Sigma N(N)}|
\bk \sigma_\Lambda, -\bk \sigma \tau \rangle \notag \\
 & \sum_{\bk_h} \sum_{\sigma_h, \tau_h}  \left(\frac{1}{\sqrt{2}}\right)^2
\langle (\bk' \sigma_\Sigma \tau_\Sigma, -\bk' \sigma' \tau', \bk_h \sigma_h \tau_h)
\notag \\
 & - (\bk' \sigma_\Sigma \tau_\Sigma, \bk_h \sigma_h \tau_h, -\bk' \sigma' \tau')|
 V_{TPE}^{\Lambda-\Sigma}  \notag \\
 &\times |(\bk \sigma_\Lambda, -\bk \sigma \tau, \bk_h \sigma_h \tau_h)
 -(\bk \sigma_\Lambda, \bk_h \sigma_h \tau_h, -\bk \sigma \tau)\rangle.
\end{align}
Explicit expressions after carrying out the spin-summation
and $\bk_h$ integration are given in Appendix. When the condition of $|\bk'|=|\bk|$
is imposed for the expression in Appendix, the expression given
by Petschauer \textit{et al.} \cite{PET17} is reproduced. Again, an additional
statistical factor of $\frac{1}{2}$ is multiplied, when it is incorporated to
the original two-body $\Lambda$N-$\Sigma$N transition interaction.

\subsection{Assignment of low-energy-constants in 3B YNN interactions}
The low energy constants in Eq. (1) were estimated in Ref. \cite{PET17}
by adopting decouplet saturation. Numerical values are
\begin{equation}
-(3b_0+b_D)=0, \hspace{1em} 2b_2+3b_4=-\frac{C^2}{\Delta} \approx -3.0
\;\mbox{GeV}^{-1},
\label{eq:LELL}
\end{equation}
where $C=\frac{3}{4}g_A$ and $\Delta$ is the average decouplet-octet mass
splitting of about 300 MeV. The coupling constants contained in Eq. (\ref{eq:LSLCa}) are
estimated as
\begin{equation}
 N_1^a=N_2^a=0,\hspace{1em}N_3^a=-\frac{4g_A^2}{3}\frac{C^2}{\Delta}\approx -6.7
\;\mbox{GeV}^{-1}.
\label{eq:num4}
\end{equation}
Similarly, those in Eq. (\ref{eq:LSLCb}) are
\begin{align}
 N_1^b&=0,\hspace{1em} N_2^b=-4\frac{8Dg_A}{9}\frac{C^2}{\Delta}\approx -10.7
 \;\mbox{GeV}^{-1},
 \notag \\ N_3^b& =-4\frac{8Dg_A}{9}\frac{C^2}{\Delta}\approx -2.7
\;\mbox{GeV}^{-1},
\label{eq:num5}
\end{align}
where $D$ is an SU(3) $D$-type coupling constant and has a relation $D+F=g_A$
with the $F$-type coupling constant.

The nucleon part of the diagram of Fig. \ref{LSdiag}(b) is the same as that in
the nucleon two-pion exchange 3NF. Namely, the diagram of Fig. \ref{LSdiag}(b)
is obtained by replacing the $\pi NN$ vertex in 3BFs by the $\pi\Lambda\Sigma$
vertex. Therefore, the corresponding coupling constants used in the nucleon
sector may be employed. Then the following estimation is possible:
\begin{align}
 N_1^b &=-D g_A 4c_1 m_\pi^2=1.6 \;\mbox{GeV}^{-1},\notag \\
 N_2^b &=2Dg_A c_3 =-6.8\;\mbox{GeV}^{-1} \notag \\
 N_3^b &= -Dg_A c_4=-3.4 \;\mbox{GeV}^{-1}.
\label{eq:numn}
\end{align}
It is reassuring to find that the corresponding numbers of Eqs. (\ref{eq:num4})
and (\ref{eq:numn}) are of the same order. In the numerical evaluations
in the next section, the values in Eqs. (\ref{eq:LELL}), (\ref{eq:num4}),
and (\ref{eq:num5}) are used. 

The diagram of Fig. \ref{LSdiag}(d) is a medium modification of the two-pion
exchange $\Lambda$N-$\Sigma$N transition by Pauli blocking.
This effect is expected to enhance the tensor component. Figure \ref{LSdiag}(f)
is a modification of the pion propagation and Figs. \ref{LSdiag}(g) and \ref{LSdiag}(h)
are regarded as a correction for the $\pi NN$ vertex in the nuclear medium. 

\section{Numerical results of YNN contributions}
Including the normal-ordered two-body interactions of the 3BFs, s.p. potentials
of the $\Lambda$ and $\Sigma$ hyperons are evaluated in SNM and PNM.
Contributions of the $\Lambda$NN-$\Lambda$NN 3BFs are first presented,
and then the effects of the $\Lambda$NN-$\Sigma$NN transition interactions
are incorporated.

\subsection{2$\pi$-exchange $\Lambda$NN-$\Lambda$NN interaction}
It is instructive to estimate the effect of the $\Lambda$NN-$\Lambda$NN 3BFs
to the $\Lambda$ s.p. potential by considering their contribution on the Hartree-Fock
level. Namely, the following summation over two nucleons in occupied states is evaluated.
\begin{align}
 & \Delta U_\Lambda (k_\Lambda) = \frac{1}{2} \sum_{\sigma_\Lambda}\frac{1}{2}
\sum_{|\bk|\le k_F, \sigma, \tau} \sum_{ |\bk'|\le k_F, \sigma', \tau'} \notag \\
 & \times  \langle \bk_\Lambda \sigma_\Lambda, \bk \sigma \tau, \bk' \sigma' \tau'|
V_{TPE}^{\Lambda NN} \notag \\
 & \times |  (\bk_\Lambda \sigma_\Lambda, \bk \sigma \tau, \bk' \sigma' \tau')
 - (\bk_\Lambda \sigma_\Lambda, \bk' \sigma' \tau', \bk \sigma \tau )\rangle .
\end{align}
When the cutoff form factor is disregarded, the result does not depend on $k_\Lambda$
and the summation can be carried out analytically to yield
\begin{align}
 & \Delta U_\Lambda = -\frac{g_A^2}{3f_0^4}\frac{1}{(2\pi)^6} F_\tau
 \frac{4\pi^2}{3}\left[\rule[0em]{0em}{1.5em} (3b_0+b_D)m_\pi^2  \right. \notag \\
 & \times \left\{ \frac{3}{2} k_F^2(2k_F^2-m_\pi^2)
 + \frac{3}{8} m_\pi^2 (8k_F^2+m_\pi^2) \log\frac{4k_F^2+m_\pi^2}{m_\pi^2}\right. \notag \\
 & \left.  -6 m_\pi k_F^3 \arctan \frac{2k_F}{m_\pi} \right\}\notag \\
 & +(2b_2+3b_4)  \left\{-\frac{1}{2} m_\pi^4(9 k_F^2 +m_\pi^2)
 \log\frac{4k_F^2+m_\pi^2}{m_\pi^2} \right. \notag \\
 &\left.\left. +\frac{2}{3}k_F^2( 2k_F^4-9m_\pi^2 k_F^2+3m_\pi^4)
 +10 m_\pi^3 k_F^3 \arctan \frac{2k_F}{m_\pi} \right\} \right].
\label{eq:delu}
\end{align}
where $F_\tau$ is a factor from isospin summation: $F_\tau=6$ in SNM
and $F_\tau=1$ in PNM.

\begin{figure}[t]
\centering
 \includegraphics[width=0.4\textwidth,clip]{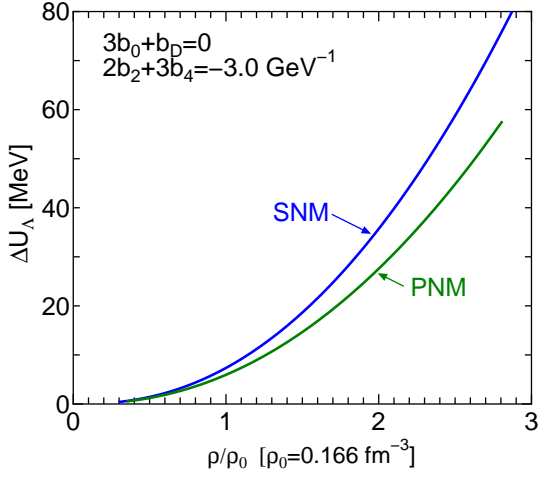}
 \caption{$k_F$ dependence of $\Delta U_\Lambda$, Eq. \ref{eq:delu}, in SNM
and in PNM. Coupling constants estimated by Petschauer \textit{et al.}
\cite{PET17} are employed.}
\label{fig:HF3BF}
\end{figure}

The $k_F$ dependence of $\Delta U_\Lambda$ is shown in Fig. \ref{fig:HF3BF}
both for SNM and PNM. Using the low-energy constants of Eq. (\ref{eq:LELL}),
the repulsive effect from the $\Lambda$NN-$\Lambda$NN interaction is about
7 MeV at the normal density in SNM and grows rapidly with increasing the density.
It can be numerically proved that the inclusion of a cutoff form factor
of the form of $\exp [-\frac{1}{36\Lambda_{3BF}^4}\{(\bk_\Lambda-\bk)^2
+(\bk_\Lambda-\bk')^2+(\bk-\bk')^2\}^2]$ modifies the result little, with the cutoff
momentum of $\Lambda_{3BF} \approx 500$ MeV.

The $\Lambda$N ladder correlation by the $G$-matrix equation is expected to
somewhat reduce the repulsive effect of the $\Lambda$NN-$\Lambda$NN interaction.
Although the 3BF in ChEFT is not applicable beyond the cutoff momentum scale,
the considerably repulsive effect without an ad hoc phenomenological
parameterization is suggestive for the properties of the $\Lambda$ hyperon
in high density nuclear matter, together with the shallow $\Lambda$ s.p.
potential presented in the previous section.

\begin{figure}[H]
\centering
 \includegraphics[width=0.4\textwidth,clip]{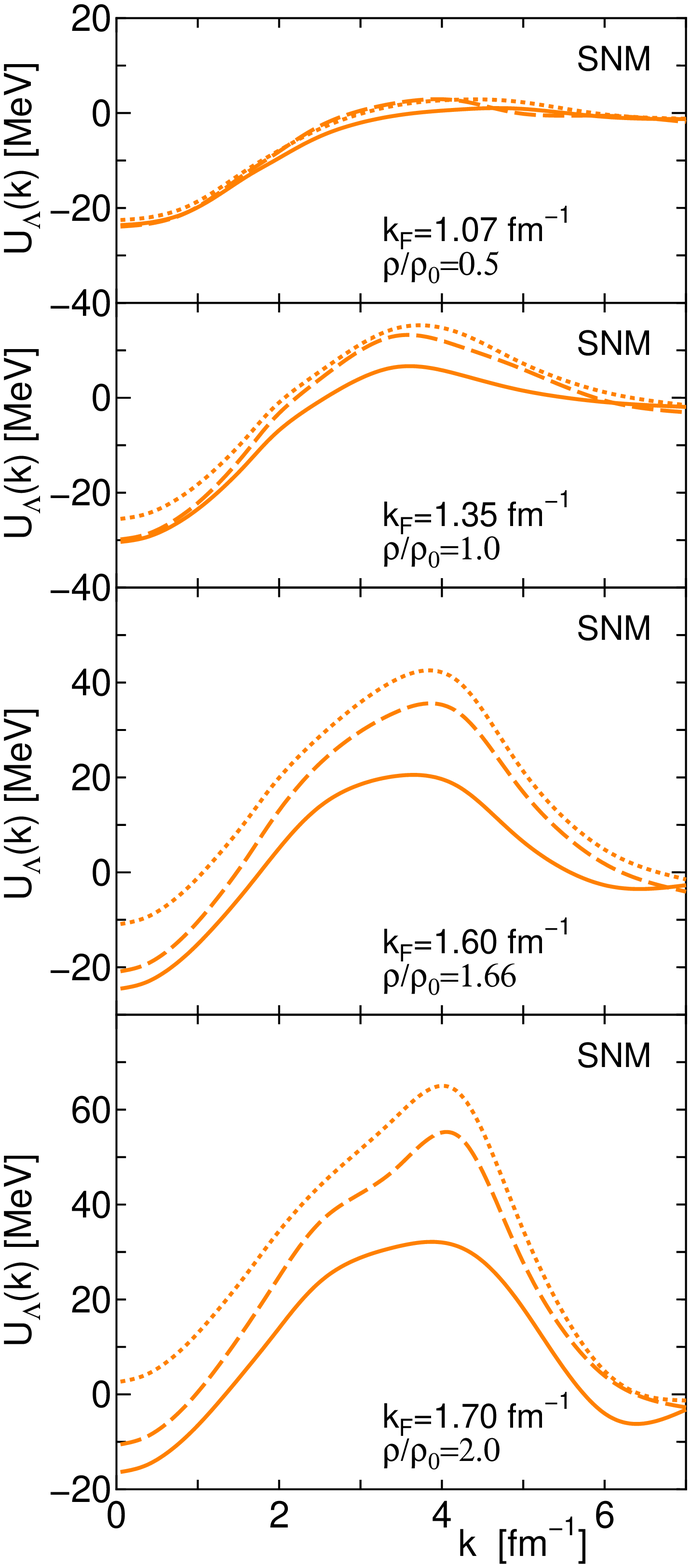}
 \caption{Real part of the $\Lambda$ s.p. potentials in SNM with the chiral
NLO YN interactions \cite{HAID13}. The solid curves are the same as in
Fig. \ref{fig:SNMR0}; that is, without the effects of 3BFs. The dotted
curves stand for the results including the effects of $\Lambda$NN-$\Lambda$NN
3BFs. The dashed curves represent the results, including the both effects
of $\Lambda$NN-$\Lambda$NN and $\Lambda$NN-$\Sigma$NN 3BFs.
}
\label{fig:SNM3BF}
\end{figure}

Now, the results of the $\Lambda$ s.p, potential obtained from the actual
$G$-matrix calculations, in which the effective two-body interaction
$V_{\Lambda N(N)}$ is incorporated,
are shown by the dotted curves in Fig. \ref{fig:SNM3BF} for SNM and
in Fig. \ref{fig:PNM3BF} for PNM. As noted earlier, stable results are not
available at high densities. The upper limit of the Fermi momentum is $k_F=1.70$
fm$^{-1}$ in SNM and $k_F^n=2.00$ fm$^{-1}$ in PNM. The $\Sigma$ s.p. potential
is not shown, because the $\Sigma$NN-$\Sigma$NN 3BFs are not considered
and the change due to the variation of the $\Lambda$ s.p. potential is very small.
The repulsive effect of the $\Lambda$NN-$\Lambda$NN interaction is about
5 MeV at the normal density and about 20 MeV at $2\rho_0$ in SNM.
The similar repulsive contribution is also obtained in PNM. Compared with
$\Delta U_\Lambda$ in Fig. \ref{fig:HF3BF}, the $G$-matrix ladder correlation
reduces the repulsion by $30\sim 40$ \%. 

In this article, the one-pion exchange contact and three-body contact terms
of the leading order YNN interactions are not considered. The estimation of their role
by Haidenbauer \textit{et al.} Ref. \cite{HMKW17} shows that the attractive contribution
can be obtained to compensate the repulsive $\Lambda$NN-$\Lambda$NN effect
by tuning the sign of the coupling constants. Before quantifying these contributions,
however, it is better to investigate the effect of the $\Lambda$NN-$\Sigma$NN
3BFs.

\begin{figure}[t]
\centering
 \includegraphics[width=0.4\textwidth,clip]{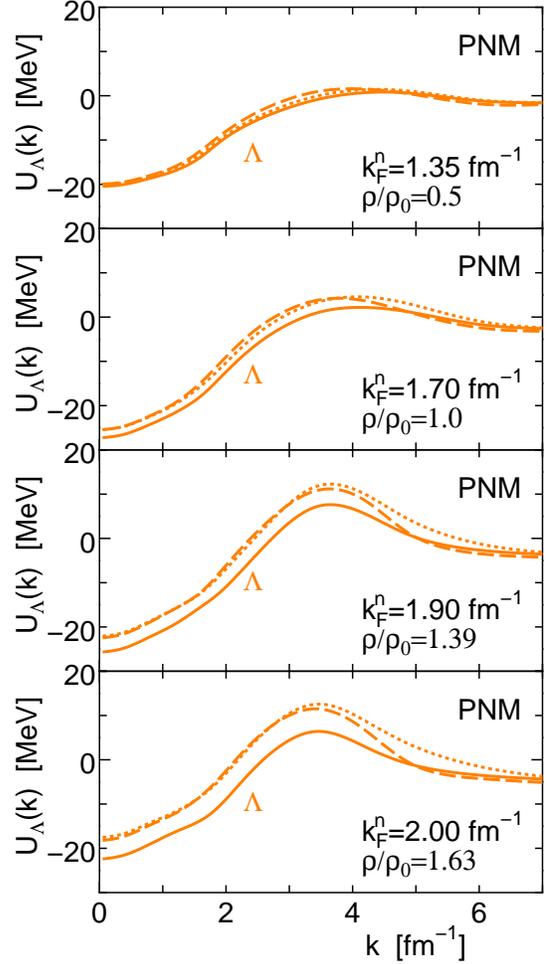}
 \caption{Same as Fig. \ref{fig:SNM3BF} but for PNM.}
\label{fig:PNM3BF}
\end{figure}

\subsection{2$\pi$-exchange $\Lambda$NN-$\Sigma$NN interaction}
Results of the $G$-matrix calculations including the effects both from
the $\Lambda$NN-$\Sigma$NN and $\Lambda$NN-$\Sigma$NN 3BFs are
presented by the dashed curves in Fig. \ref{fig:SNM3BF} for SNM and
in Fig. \ref{fig:PNM3BF} for PNM.

It is interesting to observe that the net attractive contribution from the
$\Lambda$NN-$\Sigma$NN coupling almost cancels the repulsive effect
of the $\Lambda$NN-$\Lambda$NN 3BFs at the normal density in SNM, and
the potential depth of the $\Lambda$ hyperon remains to be consistent with
experimental data. At higher densities, however, the cancellation is incomplete
and the net 3BF contribution is repulsive. In PNM, on the other hand, the effect
of the $\Lambda$NN-$\Sigma$NN 3BFs is very small, and therefore does not
weaken the repulsion from the $\Lambda$NN-$\Lambda$NN 3BFs.

\subsection{Relevance to the role of $\Lambda$ in neutron star matter}
The $\Lambda$ s.p. potential in the nuclear medium is closely related to
the possible role of the $\Lambda$ hyperons in neutron star matter.
Although the mixture of protons has to be taken into account in the realistic
study of neutron star matter, it is helpful to examine the density-dependence
of the neutron chemical potential $\mu_n$ and the $\Lambda$ s.p. potential
$U_\Lambda(0)$ at rest in PNM. Considering the $\Lambda$-neutron mass
difference $\Delta m=m_\Lambda-m_n=176.1$ MeV, the simple condition
for the emergence of the $\Lambda$ hyperon in PNM is
\begin{equation}
 U_\Lambda(0) < \mu_n -\Delta m
\end{equation}

Although the present calculations with the interactions in ChEFT are limited to
the densities below $1.63 \rho_0$, it is worth considering the implication of
the density-dependence of the present $U_\Lambda(0)$.
Figure \ref{fig:LTH} shows the density-dependence of the calculated $U_{\Lambda}(0)$
and that of $\mu_n-m_\Delta=\frac{\hbar^2}{2m_n}(k_F^n)^2 +U_n(k_F^n)-\Delta m$.
The solid and dashed curves of $U_\Lambda (0)$ are the results with and without
including YNN 3BF effects, respectively. To see the difference
of the ChEFT YN interaction from other YN potentials, two curves are included,
showing the results of the Brueckner-Hartree-Fock (BHF) calculations
by Schulze and Rijken \cite{SCH11} with employing the Nijmegen YN potentials
NSC89 \cite{NIJ89} and ESC08 \cite{RNY10}, in which YNN 3BFs are
not taken into account.

For the neutron chemical potential $\mu_n$, three cases of the ChEFT cutoff scale,
$\Lambda=450$, 550, and 600 MeV, are presented up to the density $1.63 \rho_0$,
$1.63\rho_0$, and $1.19\rho_0$, respectively.
For comparison, typical other nucleon chemical potentials are included in Fig. \ref{fig:LTH},
which are evaluated by $\mu_\Lambda=\frac{\partial \epsilon (\rho_n,\rho_\Lambda)}
{\partial \rho_\Lambda}$ at $\rho_\Lambda=0$, using the energy density
$\epsilon(\rho_n,\rho_\Lambda)$ found in the literature.
The dashed curve denotes $\mu_n$ evaluated from the parameterization of the
energy density of the model A18+$\delta v$+UIX* given in Appendix of the article
by Akmal, Pandharipande, and Ravenhall \cite{APR98} on the basis of .the variational
calculation with AV18 \cite{AV18} NN force + Urbana model IX 3BF \cite{PPCW95}.
The short dashed curve is obtained by the parameterization by Schulze and
Rijken \cite{SCH11}, which is based on the BHF calculation
with AV18 \cite{AV18} + UIX' 3BF \cite{ZH04}. The EOS corresponding to
these $\mu_n$ may not be stiff enough to support a neutron star with a mass
of 2$M_\odot$ \cite{SCH11}, but serves as a standard EOS with which various
models of the neutron matter EOS in the literature can be compared.

The curves of $U_\Lambda(0)$ with NSC98 and ESC08 cross the curve of the
standard $\mu_n$ below the neutron matter density of $2.5\rho_0$. The
onset of the $\Lambda$ emergence makes the EOS appreciably soft, with
which a heavy neutron star with a mass of $2M_\odot$ cannot be held.
It is a natural hypothesis to include strongly repulsive $\Lambda$NN 3BF
\cite{LON15,YAM14} to avoid the $\Lambda$ emergence or to make
the EOS stiff enough even under the presence of $\Lambda$ hyperons. 
The present chiral YN interactions suggest a somewhat different possibility.
The dashed curve in Fig. \ref{fig:LTH} indicates that the $\Lambda$ s.p. potential
is considerably shallow. The naive extrapolation of the dashed curve to
the higher density region may not cross the curve of the standard $\mu_n$.
The 3BFs constructed systematically in ChEFT provides an additional repulsive
effect, which can be sizable at higher densities as Fig. \ref{fig:HF3BF} denotes.
Supposing that it may not be appropriate to discuss neutron star matter at
densities beyond $3\sim 4 \rho_0$ by a solely hadronic picture,
the non-emergence of the $\Lambda$ hyperon below the density
of $3\sim 4 \rho_0$ can be a solution of the hyperon puzzle.

\begin{figure}[t]
\centering
 \includegraphics[width=0.4\textwidth,clip]{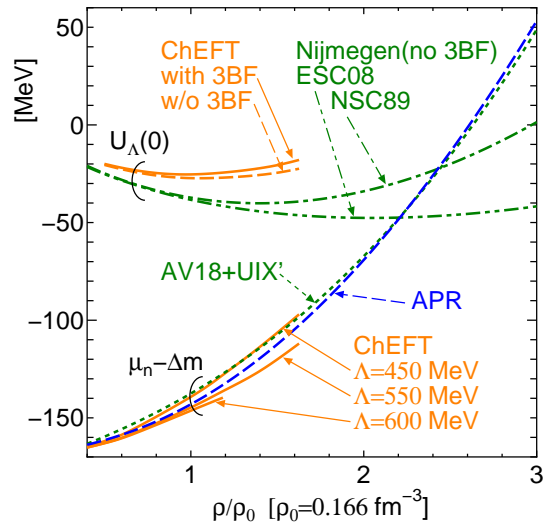}
 \caption{$k_F$ dependence of $U_\Lambda(0)$ and $\mu_n-\Delta m$
in PNM. Solid and dashed curves for the $U_\Lambda(0)$ stand for
the results with and without 3BF contribution, respectively. The dashed curve
is based on the energy density parameterized by Akmal, Pandharipande,
and Ravenhall \cite{APR98}, and the short dashed curve is evaluated by
the parameterization by Schulze and Rijken \cite{SCH11}. The dot-dashed curve and
the two-dot-dashed curve for $U_\Lambda(0)$ are calculated using the energy
densities by Schulze and Rijken \cite{SCH11}.
}
\label{fig:LTH}
\end{figure}

\section{Summary}
Single-particle potentials of the $\Lambda$ and $\Sigma$ hyperons in SNM and in PNM
are calculated in the standard LOBT framework, using YN interactions \cite{HAID13}
constructed in chiral effective field theory at the NLO level. Contributions of the
NNLO YNN 3BFs \cite{PET16} are also considered in the normal-ordering prescription.
In addition to the effects of the $\Lambda$NN-$\Lambda$NN 3BF, those of the
$\Lambda$NN-$\Sigma$NN transition 3BF are incorporated. The $\Lambda$N-$\Sigma$N
coupling is known to be important because of the tensor component
from the pion exchange. Compared with other YN potentials, the ChEFT YN interaction
bears particularly strong $\Lambda$N-$\Sigma$N coupling. The $\Lambda$ s.p. potential
is repulsive when the $\Lambda$N-$\Sigma$N coupling is switched off.
The $\Lambda$ s.p. potential of about $-30$ MeV at the normal density, which is
consistent with the empirical data, originates from the $\Lambda$N-$\Sigma$N coupling.
The $\Sigma$ potential is weakly attractive at the low densities in SNM, but becomes
repulsive with increasing the density. This feature is consistent with the empirical data.
The $\Sigma$ potential hight in SNM and the $\Sigma^-$ potential hight in PNM grow
rapidly as the density increases.  

The salient feature of the NLO YN interaction is that the $\Lambda$ s.p. potential
becomes shallower at greater than normal density. This behavior is paved by the
density dependence of the contribution of the $\Lambda$N-$\Lambda$N interaction
in the $^3$S$_1$ channel. In addition, the $\Lambda$N-$\Sigma$N coupling bringing
about the substantial attraction tends to be suppressed by Pauli blocking
when the Fermi momentum becomes larger. The situation is same in PNM,
which is more relevant to neutron star matter.

It has been expected that the $\Lambda$ hyperons appear in high-density neutron
star matter to bypass the neutron chemical potential, and therefore the equation
of state of PNM becomes soft. The recent observation of twice the solar-mass neutron
stars has imposed a constraint for the stiffness of the EOS, with which
naive emergence of hyperons seems to be excluded. This contradictory situation is
called hyperon puzzle. The character of the $\Lambda$ s.p. potential predicted by
the NLO ChEFT suggests
that the $\Lambda$ hyperon is energetically not favored in neutron star matter at
higher densities. Because ChEFT is effective low-energy theory, the
cutoff scale being of the order of 500 MeV, it is not adequate to discuss the region of
the density greater than $2\sim 2.5$ times normal density. Nevertheless,
the behavior of the $\Lambda$ s.p. potential below the limit is suggestive, and the
ChEFT YN interaction provides a possibility to resolve the hyperon puzzle.

The effects of the two-pion exchange $\Lambda$NN-$\Lambda$NN 3B interactions
make the $\Lambda$ s.p. potential less attractive. To be consistent with the empirical
data, the attraction which compensates this repulsive contribution is required.
It is conceivable to include the $\Lambda$NN-$\Lambda$NN contact terms,
as was practiced by Haidenbauer \textit{et al.} \cite{HMKW17}, and tune the coupling
constants. In this article, before introducing those contact terms, contributions from
the $\Lambda$NN-$\Sigma$NN transition 3BF are investigated. It is demonstrated
that the effects of the effective $\Lambda$N-$\Sigma$N interactions from
the $\Lambda$NN-$\Sigma$NN processes serve to restore the $\Lambda$ potential
depth of about 30 MeV at the normal density in SNM. At the densities greater than
the normal, the net contribution from the YNN interaction is repulsive
of the order of $5\sim 10$ MeV, which supports to disfavor the emergence of
the $\Lambda$ hyperon in high-density neutron star matter.

It is known in the non-strange sector that higher orders in ChEFT beyond NLO are
important. In the future, the inclusion of higher order terms is necessary also in the
strangeness sector, although at present it is not meaningful because of the
insufficiency of experimental data on the amount as well as the accuracy.
YNN 3BFs of $K$ and $\eta$ meson exchange processes are better to be
considered. In parallel with these developments, it is also important to examine
the properties of the ChEFT YN interactions by investigating experimental data of
hyper nuclei, which will be provided by the on-going and future experiments. 

\acknowledgments
The author is grateful to J. Haidenbauer for providing him with the computational
code of hypron-nucleon interactions of ChEFT and for valuable discussions.
This work is supported by JSPS KAKENHI Grant (No. JP15H00837 and No. JP16K17698).

\newpage

\appendix
\begin{widetext}
\bigskip
{\footnotesize\bf APPENDIX: EFFECTIVE TWO-BODY FORCES FROM THE 3NF IN CHIRAL EFFECTIVE FIELD THEORY}
\subsection{Effective 2B interaction from the $\Lambda$NN-$\Lambda$NN interaction}
The evaluation of Eq. (2) becomes
\begin{align}
& \frac{g_A^2}{3f_0^4} C_\tau^{\Lambda\Lambda} (3b_0+b_D) m_\pi^2  \sum_{\ell=0}^\infty
 \hat{\ell} P_\ell(\cos\theta) \left[ \cos\theta \;Q_{W0}^\ell(k',k)
  - \{Q_{X1}^\ell(k,k')+Q_{X1}^\ell(k',k)-\delta_{\ell 0}\frac{1}{2}(F_0(k')+F_0(k))\}\right]\notag \\
& -\frac{g_A^2}{3f_0^4} 3 (2b_2+3b_4) \left\{ \frac{1}{8}\rho_0
-\frac{1}{4}(2k'^2+k^2+3m_\pi^2)F_0(k')
 -\frac{1}{4}(k'^2+2k^2+3m_\pi^2)F_0(k) \right. \notag \\
 & +\frac{1}{4}(k'^2 F_2(k')+k_1^2 F_2(k)) + k'k P_1(\cos\theta)
 \left[F_0(k')+F_0(k)-\frac{1}{2}(F_1(k')+F_1(k)))\right] \notag \\
 & \left. + \frac{1}{4} \left(2m_\pi^2 +k'^2 +k^2 -2k' k \cos\theta
\right)^2 \frac{1}{k'k}
 \sum_\ell \hat{\ell} P_\ell (\cos\theta) Q_{W0}^\ell (k',k) \right\} \notag \\
&+\frac{g_A^2}{3f_0^4} 3 (i\bfsigma_N\cdot (\bk'\times\bk))(3b_0+b_D)m_\pi^2 \frac{1}{k'k}
 \sum_{\ell=0}^\infty \left\{ \hat{\ell} P_\ell(\cos\theta)
Q_{W0}^\ell(k',k) \right. \notag \\
 &\left. + P_{\ell}'(\cos\theta) (Q_{W1}^{\ell, \ell+1}(k,k') +Q_{W1}^{\ell, \ell+1}(k',k)
 -Q_{W1}^{\ell, \ell-1}(k,k')-Q_{W1}^{\ell, \ell-1}(k',k))  \right\}\notag \\
&-\frac{g_A^2}{3f_0^4} 3(2b_2+3b_4) (i\bfsigma_N\cdot (\bk'\times\bk))
 \left[\frac{1}{2}(F_0(k')+F_0(k)-F_1(k')-F_1(k))
 - \frac{((\bk'-\bk)^2+2m_\pi^2)}{2k'k} \sum_{\ell=0}^\infty \right. \notag \\
  &\left. \times \{ \hat{\ell} P_\ell(\cos\theta)Q_{W0}^\ell(k',k)
 +P_{\ell}'(\cos\theta) (Q_{W1}^{\ell, \ell+1}(k,k') +Q_{W1}^{\ell, \ell+1}(k',k)
 -Q_{W1}^{\ell, \ell-1}(k,k')-Q_{W1}^{\ell, \ell-1}(k',k)) \} \right],
\end{align}
where $P_\ell(\cos\theta)$ is a Legendre function and $\theta$ is the angle
between $\bk'$ and $\bk$. Definitions of $F_0$, $F_1$, $F_2$, $Q_{W0}^\ell$,
$Q_{W1}^\ell$, and $Q_{W1}^{\ell, \ell+1}$ are given in Ref. [24].
The constant $C_\tau^{\Lambda\Lambda}$ comes from the isospin
summation: $C_\tau^{\Lambda\Lambda}=3\delta_{\tau' \tau}$ in SNM and
$C_\tau^{\Lambda\Lambda}=\{1\delta_{\tau' -1/2}\delta_{\tau -1/2}+2\delta_{\tau 1/2}\delta_{\tau 1/2}\}$
in PNM. If the condition $|\bk|=|\bk'|$
is applied for the initial and final momenta, $\bk$ and $\bk'$, the expression
given by Petschauer \textit{et al.} [31] is recovered. 
The term including $(i\bfsigma_2\cdot (\bk'\times\bk))$ gives normal
and anti-symmetric spin-orbit components:
\begin{equation}
 i\bfsigma_N\cdot (\bk'\times\bk)= -i\frac{1}{2}(\bfsigma_\Lambda+\bfsigma_N)\cdot
(\bk\times\bk')+i\frac{1}{2}(\bfsigma_\Lambda-\bfsigma_N)\cdot(\bk\times\bk').
\end{equation}

The partial wave decomposition is obtained by operating the following integration for
the above expression:
\begin{itemize}
\setlength{\itemsep}{5pt}
\setlength{\parskip}{5pt}
 \item[(1)] the central component,
 $\frac{1}{2}\int_{-1}^1 d\cos\theta  P_\ell (\cos\theta)$;
 \item[(2)] the spin-orbit component,
 $\frac{1}{2}\int_{-1}^1 d\cos\theta \;k'k \delta_{\ell'\ell} \delta_{SS'}
\delta_{S 1} (-1)^{1}\frac{\ell(\ell+1)+2-J(J+1)}{2\hat{\ell}}
 \left\{ P_{\ell-1}(\cos \theta) -P_{\ell+1}(\cos \theta) \right\}$ for the coefficient of
$\frac{1}{2}i(\bfsigma_\Lambda +\bfsigma_N)\cdot (\bk\times \bk')$;
 \item[(3)] the antisymmetric spin-orbit component,
 $\frac{1}{2}\int_{-1}^1 d\cos\theta \;k'k \delta_{S\ne S'}\delta_{\ell'\ell}
\delta_{\ell J} \sqrt{\ell(\ell+1)} \frac{1}{\hat{\ell}}
 \{P_{\ell-1}(\cos \theta)-P_{\ell+1}(\cos \theta) \}$ for the coefficient of
$\frac{1}{2}i(\bfsigma_\Lambda -\bfsigma_N)\cdot (\bk\times \bk')$.
\end{itemize}

\subsection{Effective 2B interaction from the $\Lambda$NN-$\Sigma$NN interaction}
\subsubsection{$V_{TPE,a}^{\Lambda-\Sigma}$ in Eq. (4)}
In the following expressions, a factor $2\sqrt{3}$ for the isospin $\frac{1}{2}$ base is included
in the coupling constants. For PNM, an additional factor is multiplied:
$0$ for $\Lambda n\rightarrow \Sigma^0 n$, $\frac{1}{2}$ for  $\Lambda n\rightarrow \Sigma^- p$,
$1$ for  $\Lambda p\rightarrow \Sigma^0 p$, and $\frac{1}{2}$ for $\Lambda p\rightarrow \Sigma^+ n$.
The central component from the evaluation of Eq. (6) becomes 
\begin{align}
% & \fbox{central} \notag \\
 & -\frac{1}{4f_0^4} N_1^{a} \; \sum_{\ell=0}^\infty
 \hat{\ell} P_\ell(\cos\theta) \left\{ \cos\theta Q_{W0}^\ell(k',k)
 -Q_{X1}^\ell(k,k')-Q_{X1}^\ell(k',k)+\delta_{\ell 0}\frac{1}{2}(F_0(k')+F_0(k))
 + Q_{W2}^\ell (k',k) \right\}  \notag \\
 & +\frac{1}{4f_0^4}N_2^{a} \left\{ \frac{1}{8}\rho_0
 -\frac{1}{4}(2k'^2+k^2+3m_\pi^2)F_0(k')
 -\frac{1}{4}(k'^2+2k^2+3m_\pi^2)F_0(k) \right.\notag \\
 & +\frac{1}{4}(k'^2 F_2(k')+k^2 F_2(k)) + k'k P_1(\cos\theta)
 \left[F_0(k')+F_0(k)-\frac{1}{2}(F_1(k')+F_1(k))\right] \notag \\
 & \left. +\frac{1}{4} \left(2m_\pi^2 +k'^2 +k^2 -2k'k\cos\theta \right)^2 \frac{1}{k'k}
 \sum_\ell \hat{\ell} P_\ell (\cos\theta) Q_{W0}^\ell (k',k)\right\}\notag \\
 & +N_3^{a} \frac{1}{4f_0^4}\frac{1}{3}( \bfsigma_Y \cdot \bfsigma_N)
 \left[\frac{1}{4}(3k'^2+k^2)F_0(k')+\frac{1}{4}(k'^2+3k^2)F_0(k)-\frac{1}{2}(k'^2F_1(k')+k^2F_1(k))
 \right. \notag \\
 &-\frac{1}{2}k'k\cos\theta \{2(F_0(k')+F_0(k))-(F_1(k')+F_1(k'))\}\notag \\
 &\left. -\frac{1}{4}(k'^2+k^2-2k'k\cos\theta)(k'^2+k^2+4m_\pi^2-2k'k\cos\theta)
 \sum_\ell \hat{\ell}P_\ell (\cos\theta) \frac{1}{k'k}Q_{W0}^\ell (k',k)\right].
\end{align}

The spin-orbit and antisymmetric spin-orbit terms are
\begin{align}
&-\frac{1}{4f_0^4} N_1^{a}(i\bfsigma_N\cdot (\bk'\times\bk)) \frac{1}{k'k}
 \sum_{\ell=0}^\infty \left\{ \hat{\ell} P_\ell(\cos\theta)
Q_{W0}^\ell(k',k) \right. \notag \\
 &\left. + P_{\ell}'(\cos\theta)(Q_{W1}^{\ell,\ell+1}(k,k')+Q_{W1}^{\ell,\ell+1}(k',k)
 -Q_{W1}^{\ell, \ell-1}(k,k')-Q_{W1}^{\ell, \ell-1}(k',k))  \right\}\notag \\
&+\frac{1}{4f_0^4} N_2^{a} (i\bfsigma_N\cdot (\bk'\times\bk))
 \left[\frac{1}{2}(F_0(k')+F_0(k)-F_1(k')-F_1(k))
 - \frac{((\bk'-\bk)^2+2m_\pi^2)}{2k'k} \sum_{\ell=0}^\infty \right. \notag \\
  &\left. \times \{ \hat{\ell} P_\ell(\cos\theta)Q_{W0}^\ell(k',k)
 +P_{\ell}'(\cos\theta) (Q_{W1}^{\ell, \ell+1}(k,k') +Q_{W1}^{\ell, \ell+1}(k',k)
 -Q_{W1}^{\ell, \ell-1}(k,k')-Q_{W1}^{\ell, \ell-1}(k',k)) \} \right] \notag \\
 &+\frac{1}{4f_0^4} N_3^{a} (i\bfsigma_Y \cdot (\bk'\times \bk))
\left[ \frac{1}{2}(F_0(k')+F_0(k)- F_1(k')- F_1(k ))
 -\frac{((\bk'-\bk)^2+2m_\pi^2)}{2k'k}\right. \notag \\
 & \left.  \times  \sum_\ell \left\{\hat{\ell} P_\ell (\cos\theta) Q_{W0}^\ell (k',k)
  +P_{\ell}'(\cos\theta)( k' Q_{S1}^{\ell+1,\ell}(k',k)+kQ_{S1}^{\ell+1,\ell}(k,k')
 -k'Q_{S1}^{\ell-1,\ell}(k',k) -kQ_{S1}^{\ell-1,\ell} (k,k')) \right\}\rule{0mm}{6mm} \right].
\end{align}

Finally, the tensor part is
\begin{align}
 & +N_3^{a}\frac{1}{4f_0^4}\left[ \frac{1}{3}\sum_\ell \hat{\ell} P_\ell (\cos\theta)
 \{-k^2 S_{12}(\bk',\bk') -k'^2 S_{12}(\bk,\bk)+2(\bk'\cdot\bk) S_{12}(\bk',\bk)\}
 \frac{1}{k'k} Q_{W0}^\ell (k',k) \right.  \notag \\
 &+ \frac{1}{3} 4\pi \sum_{j' j k} Q_{W1}^{j' j} (k',k) \sqrt{\hat{j'}\hat{j}}
 (-1)^{j'} (j' 0 10|j 0) (10 j 0|k0) \sixj{1}{2}{1}{j'}{j}{k} \notag \\
 & \times \left\{ (k'^2+k^2)  
 ([\bfsigma_1\times \bfsigma_2]^2\cdot[Y_{j'}(\hat{\bk'})
\times Y_{k}(\hat{\bk})]_{\mu}^{2})
 +2k' k \sum_{k'k''} \sqrt{\frac{\hat{j'}\hat{k}}{3\cdot 5}} (-1)^{k'+k''}
  \sixj{j'}{k}{2}{k'}{k''}{1}\right. \notag \\
 & \left. \hspace{8em}\times (10 j' 0|k''0) (10 k 0|k'0)
  ([\bfsigma_1\times \bfsigma_2]^2 \cdot[Y_{k''}(\hat{\bk'})
\times Y_{k'}(\hat{\bk})]^{2})\right\}  \notag \\
 & + \{\bk'\leftrightarrow \bk\}  \notag \\
 &+ \frac{1}{3} \sum_{\ell=0}^\infty  \hat{\ell} P_\ell(\cos\widehat{\bk'\bk})
 \{Q_{X1}^\ell(k,k')+Q_{X1}^\ell(k',k)
-\delta_{\ell 0}\frac{1}{2}(F_0(k')+F_0(k))\}
 \frac{1}{3} S_{12}(\bk'-\bk,\bk'-\bk)  \notag \\
 & +\sqrt{\frac{7}{3}}4\pi \sum_{j' j k} Q_{W1}^{j' j} (k',k)
 \sqrt{\hat{j'}\hat{j}} (-1)^{j'}
  (j' 0 10|j 0) (10 j 0|k0)\sixj{1}{2}{1}{j'}{j}{k}  \notag \\
 & \times \left[ (1010|20) k'^2 \sum_{J}
 (-1)^{j'+k}\sqrt{\hat{J}\hat{j'}}\sixj{2}{j'}{J}{k}{2}{2}
 (20j'0|J0)([\bfsigma_1\times \bfsigma_2]^2
 \cdot[Y_J(\hat{\bk'})\times Y_{k}(\hat{\bk})]^2)\right.\notag \\
 & \hspace{1em}+ (1010|20) k^2 \sum_{J} (-1)^{j'+k}
\sqrt{\hat{J}\hat{k}}\sixj{2}{k}{J}{j'}{2}{2}
 (20k 0|J0)([\bfsigma_1\times \bfsigma_2]^2
 \cdot[Y_{j'}(\hat{\bk'})\times Y_{J}(\hat{\bk})]^2)\notag \\
 &  \left. \hspace{1em}  - 2k'k\sum_{J_{12}J_{34}}
 \sqrt{\hat{J_{12}}\hat{J_{34}}\hat{j'}\hat{k}}
\ninj{1}{j'}{J_{12}}{1}{k}{J_{34}}{2}{2}{2}
 (10j' 0|J_{12}0)(10k 0|J_{34}0)
 ([\bfsigma_1\times\bfsigma_2]^2\cdot [Y_{J_{12}}(\hat{\bk'})
 \times Y_{J_{34}} (\hat{\bk})]^2)
 \rule{0mm}{5mm}\right] \notag\\
  & +\{ \bk' \leftrightarrow \bk \} \notag \\
% & \mbox{(6)}  \notag \\
 & -\frac{1}{3}\sum_{\ell=0}^\infty \hat{\ell} P_\ell (\cos \theta) Q_{W2}^\ell (k',k)
 \frac{1}{3}S_{12}(\bk'-\bk, \bk'-\bk)  \notag \\
 & -\frac{1}{3} 4\pi \sum_{j' j} Q_{W2}^{j' j} (k',k) \frac{1}{5} \sqrt{\frac{2\hat{j'}\hat{j}}{3}}
 (j' 0 j 0|20) \left[ \rule{0mm}{8mm} (k'^2+k^2) 
 ([\bfsigma_1\times \bfsigma_2]^2 \cdot[Y_{j'}(\hat{\bk'})\times Y_{j}(\hat{\bk})]^2)\right. \notag \\
 & \left. + 2 k' k \sum_{k'k}  \sqrt{\frac{\hat{j'}\hat{j}}{15}} (-1)^{k+k'}
   \sixj{j'}{j}{2}{k}{k'}{1} (10 j' 0|k'0) (10 j 0|k0) 
 ([\bfsigma_1\times \bfsigma_2]^2 \cdot [Y_{k'}(\hat{\bk'})\times Y_{k}(\hat{\bk})]^2)
 \right]  \notag \\
 & -\sqrt{\frac{7}{3}} 4\pi \sum_{j' j} Q_{W2}^{j'j} (k',k) \frac{1}{5}
  \sqrt{\frac{2\hat{j'}\hat{j}}{3}}(j' 0 j 0|20) \left[ \rule{0mm}{8mm}(1010|20) \right. \notag \\
 & \times \left(k'^2 \sum_{k} (-1)^{j'+j} \sqrt{\hat{k}\hat{j'}} \sixj{2}{j'}{k}{j}{2}{2} (20 j' 0|k0)
 ([\bfsigma_1\times \bfsigma_2]^2 \cdot[Y_k(\hat{\bk'})\times Y_j (\hat{\bk})]^2) \right.  \notag \\
 & \left. +k^2 \sum_{k} (-1)^{j'+j} \sqrt{\hat{k}\hat{j}} \sixj{2}{j}{k}{j'}{2}{2} (20 j 0|k0)
 ([\bfsigma_1\times \bfsigma_2]^2 \cdot[Y_{j'}(\hat{\bk'})\times Y_k(\hat{\bk})]^2) \right)  \notag \\
 & \left. \left. -2k'k  \sum_{k'k} \sqrt{\hat{k'}\hat{k}\hat{j'}\hat{j}}
 \ninj{1}{j'}{k'}{1}{j}{k}{2}{2}{2}  (10 j' 0|k'0) (10 j 0|k0)
  ([\bfsigma_1\times \bfsigma_2]^2\cdot
 [Y_{k'}(\hat{\bk'})\times Y_{k}(\hat{\bk})]^2)
 \right] \rule{0mm}{10mm} \right].
\end{align}

\subsubsection{$V_{TPE,b}^{\Lambda-\Sigma}$ in Eq. (5)}
Because an isospin structure is complicated in PNM, only the results in SNM are presented.
The central component in SNM from the evaluation of Eq. (6) becomes 
\begin{align}
% & \fbox{central} \notag \\
 & \frac{\rho_0}{4f_0^4}  \frac{1}{3} (\bfsigma_Y\cdot\bfsigma_N) \sum_{\ell=0}^\infty
 \hat{\ell} P_\ell(\cos\theta) \left\{ -N_2^b \delta_{\ell 0} +(N_1^b +2N_2^b m_\pi^2)\frac{1}{2k'k} Q_{\ell}(z)
  +(N_1^b +N_2^b m_\pi^2)m_\pi^2 \frac{1}{(2k'k)^2} Q_{\ell}'(z) \right\}  \notag \\
 & + \frac{1}{4f_0^4}  \frac{1}{3} (\bfsigma_Y\cdot\bfsigma_N) \sum_{\ell=0}^\infty
 \hat{\ell} P_\ell(\cos\theta) \frac{1}{2k'k} Q_\ell (z) \left[\rule{0mm}{6mm} -N_1^b\{ (k'^2-k'k\cos\theta)(f_0(k')-F_1(k'))
 + (k^2+k'k\cos\theta)(F_0(k)-F_1(k))\} \right. \notag \\
 & +2N_3^b (k'^2+k^2-2k'k\cos\theta)\left\{ \frac{1}{\rho_0} -m_\pi^2 (F_0(k')+F_0(k))\right\}
 + (N_2^b-2N_3^b) \left\{ (k'^2-k'k\cos\theta)^2)(F_0(k')-2F_1(k')+F_3(k'
))) \right. \notag \\
  & \left. \left. + (k^2-k'k\cos\theta)^2)(F_0(k)-2F_1(k)+F_3(k)))+\frac{1}{3}(k'^2+k^2-2k'k\cos\theta)(k'^2(F_2(k')-F_3(k'))
 +k^2(F_2(k)-F_3(k))\right\}\right],
\end{align}
where $Q_\ell$ is the second kind Legendre function and $z\equiv \frac{k'^2+k^2+m_\pi^2}{2k'k}$.

There is no normal spin-orbit term. The antisymmetric spin-orbit terms in SNM are
\begin{align}
 & \frac{1}{4f_0^4}\frac{1}{2} (\bfsigma_\Lambda \times \bfsigma_N)\cdot(\bk' \times \bk)
 \sum_{\ell=0}^\infty \hat{\ell} P_\ell(\cos\theta) \frac{1}{2k'k}Q_\ell(z) [-N_1^b (F_0(k')-F_1(k')-F_0(k)+F_1(k)) \notag\\
 & +(N_2^b-2N_3^b)\{(k'^2-k'k\cos\theta)(F_0(k')-2F_1(k')+F_2(k'))-(k^2-k'k \cos\theta)(F_0(k)-2F_1(k)+F_3(k))\} ].
\end{align}
The partial wave decomposition is obtained by applying the following integration for the coefficient of
$\frac{1}{2} (\bfsigma_\Lambda \times \bfsigma_N)\cdot(\bk' \times \bk)$:
$\frac{1}{2}\int_{-1}^1 d\cos\theta \;k'k \delta_{S\ne S'}\delta_{\ell'\ell}
\delta_{\ell J} \sqrt{\ell(\ell+1)} \frac{1}{\hat{\ell}}(-1)^S
 \{P_{\ell-1}(\cos \theta)-P_{\ell+1}(\cos \theta) \}$ for the coefficient of
$\frac{1}{2}i(\bfsigma_\Lambda -\bfsigma_N)\cdot (\bk\times \bk')$.

Finally, the tensor part in SNM is
\begin{align}
 & -\frac{\rho_0}{4f_0^4} \frac{1}{3} S_{12}(\bk'-\bk,\bk'-\bk) \sum_{\ell=0}^\infty \hat{\ell} P_\ell(\cos\theta)
 \frac{1}{2k'k} \left\{N_2^b Q_\ell(z)+(N_1^b +N_2^b m_{\pi}^2) \frac{1}{2k'k}  Q_{\ell}' (z) \right\} \notag \\
  & +\frac{1}{4f_0^4}\sum_{\ell=0}^\infty \hat{\ell} P_\ell(\cos\theta)\frac{1}{2k'k} Q_\ell(z) \left[
 -N_1^{b} \left\{ \frac{1}{3}S_{12}(\bk'-\bk,\bk')(F_0(k')-F_1(k'))
   -\frac{1}{3}S_{12}(\bk'-\bk,\bk)(F_0(k)-F_1(k)) \rule{0mm}{6mm}\right\} \right. \notag \\
  & +2N_3^{b} \frac{1}{3}S_{12}(\bk'-\bk,\bk'-\bk)\left\{\frac{1}{2}\rho_0- m_\pi^2(F_0(k')+F_0(k))\right\}\notag \\
  & + (N_2^{b}-2N_3^{b}) \left\{\rule{0mm}{4mm}\frac{1}{3}S_{12}(\bk'-\bk,\bk')
(k'^2-k'k\cos\theta)(F_0(k')-2F_1(k')+F_3(k'))    \right. \notag \\
 &\hspace{6em} -\frac{1}{3}S_{12}(\bk'-\bk,\bk)(k^2-k'k\cos\theta)(F_0(k)-2F_1(k)+F_3(k))  \notag \\
 &\hspace{6em} \left. \left. +\frac{1}{3}S_{12}(\bk'-\bk,\bk'-\bk)
 \frac{1}{3}(k'^2 (F_2(k')-F_3(k')) +k^2 (F_2(k)-F_3(k)) )\right\} \right].
\end{align}
The partial-wave decomposition of the tensor term can be found in Ref. \cite{FUJI00}.
\bigskip
\end{widetext}

\end{document}